\documentclass[prd,twocolumn]{revtex4-1}

\newcommand{\eq}[1]{\begin{align}#1\end{align}}
\newcommand{\pert}[2]{
{}^{(#1)}\hspace{-0.5mm}#2
}

\def\Asc{\mathcal{A}}

\def\Psc{\mathcal{P}}

\def\PhiP{\pert{i}\Phi^{\Psc}_\ell}
\def\PhiPone{\pert{i}\Phi^{\Psc}_\textit{1}}
\def\PsiP{\pert{i}\Psi^\Psc_\ell}

\def\PsiPHATone{\pert{i}{\hat\Psi^\Psc_\textit{1}}}
\def\PhiA{\pert{i}\Phi^\Asc_\ell}
\def\PhiAone{\pert{i}\Phi^\Asc_{\textit{1}}}
\def\PsiAone{\pert{i}\Psi^\Asc_{\textit{1}}}
\def\PsiA{\pert{i}\Psi^\Asc_\ell}

\def\VPG{V_{G\, \ell}^\Psc} 
\def\VPM{V_{M\, \ell}^\Psc} 
\def\VPS{V_{MG\, \ell}^\Psc} 

\def\VAG{V_{G\, \ell}^\Asc} 
\def\VAM{V_{M\, \ell}^\Asc} 
\def\VAMone{V_{M\, \textit{1}}^\Asc} 
\def\VAS{V_{MG\, \ell}^\Asc} 

\def\VPGHAT{\hat V_{G\, \ell}^\Psc} 
\def\VPSHAT{\hat V_{MG\, \ell}^\Psc} 
\def\VPSHATone{\hat V_{MG\, \textit{1}}^\Psc} 

\def\SPG{\pert{i}{\tilde S}_{G\, \ell}^\Psc} 
\def\SPM{\pert{i}{\tilde S}_{M\, \ell}^\Psc} 

\def\SPMHAT{\pert{i}{\hat S}_{M\, \ell}^\Psc} 

\def\SAG{\pert{i}{\tilde S}_{G\, \ell}^\Asc} 
\def\SAM{\pert{i}{\tilde S}_{M\, \ell}^\Asc} 
\def\SAMone{\pert{i}{\tilde S}_{M\, \textit{1}}^\Asc} 

\def\anon{\pert{i}\alpha_\ell}
\def\bnon{\pert{i}\beta_\ell}
\def\gnon{\pert{i}\gamma_\ell}
\def\lnon{\pert{i}\lambda_\ell}
\def\knon{\pert{i}\kappa_\ell}

\def\anonone{\pert{i}\alpha_\textit{1}}
\def\bnonone{\pert{i}\beta_\textit{1}}
\def\gnonone{\pert{i}\gamma_\textit{1}}
\def\lnonone{\pert{i}\lambda_\textit{1}}
\def\knonone{\pert{i}\kappa_\textit{1}}

\usepackage{amssymb, amsmath}
\usepackage{enumitem}
\usepackage{caption}
\usepackage{mathtools}
\usepackage{color}
\usepackage{natbib}
\usepackage{natbib}
\usepackage{hyperref}

\begin{document}

\title{Nonlinear perturbations of Reissner--Nordstr\"om black holes}
\author{Mieszko Rutkowski}
\affiliation{Marian Smoluchowski Institute of Physics, Jagiellonian University, 30-348 Krak\'ow, Poland}
\email{mieszko.rutkowski@doctoral.uj.edu.pl}

\begin{abstract}
We develop a nonlinear perturbation theory of Reissner--Nordstr\"om black holes. We show that, at each perturbation level, Einstein--Maxwell equations can be reduced to four inhomogeneous wave equations, two for polar and two for axial sector. Gravitational part of these equations is similar to Regge--Wheeler and Zerilli equations with source and additional coupling to the electromagnetic sector. We construct solutions to the inhomogeneous part of wave equations in terms of sources for Einstein--Maxwell equations. We discuss $\ell=0$ and  $\ell=1$ cases separately.
\end{abstract}

\maketitle

\section{Introduction}
\label{sect:1}
Perturbative methods play an important role in General Relativity. They find application to stability analysis, gravitational radiation, cosmology, rotating stars, accretion disc, self--force, etc. Sometimes linear analysis give sufficient insight into physical phenomena, but sometimes going beyond linear order can change qualitatively linear predictions (e.g.\ Bizo\'n--Rostworowski conjecture of instability of anti-de Sitter spacetime \citep{BR11}). In the following paper, we study nonlinear perturbations of spherically symmetric solutions to Maxwell--Einstein equations.
Linear perturbation theory of Schwarzschild solution was formulated by \citet{RW57} and \citet{Z70a} and then generalised to Reissner--Nordstr\"om black hole by \citet{Z74} (see also \citep{M74}, \citep{M75}, \citep{B79}, \citep{KI04}, \footnote{\citep{Z74} contained some mistakes, later corrected by other authors}). Linear perturbations of Reissner--Nordstr\"om have also been recently discussed in context of \textit{strong cosmic censorship} conjecture (see {\citep{D18}, \citep{C18}}). Taking into account higher--order perturbation terms makes the computations significantly more difficult: equations at each order beyond linear include all the previous--order terms. This issue  has been treated by some authors - e.g.\ second order perturbations of Schwarzschild were studied by \citet{KT76,GP00,G00,NJ07,B09}. Recently, \citet{AR17} provided a robust framework to deal with nonlinear (in principle of any order) gravitational perturbations of spherically symmetric spacetimes. Present article is an extension of \citep{AR17} to both gravitational and electromagnetic nonlinear perturbations of Reissner--Nordstr\"om black holes.

Our approach is based on assumptions similar to those from \cite{AR17}. We rewrite them explicitly here, since there are some differences:
\begin{enumerate}
\item At each perturbation level, there are four master scalar variables, two in polar and two in axial sector. In each sector, they fulfil a system of two linearly--coupled inhomogeneous (homogeneous at the linear order) wave equations with potentials.
\item At each perturbation level, Regge--Wheeler variables and electromagnetic tensor components are linear combinations of master scalar variables from the suitable sector and their derivatives up to the second order. At the nonlinear orders, one needs to include additional functions to fulfil Maxwell--Einstein equations. 
\item At the linear level, relations from the previous point can be inverted to express master scalars as combinations of RW variables and electromagnetic tensor components. At the nonlinear level, we take the same expressions for the master scalar functions.
\end{enumerate}

In our considerations, we restrict ourselves to axially--symmetric perturbations only (going beyond axial symmetry is a straightforward procedure, that conceptually adds little to this paper). During calculations, we stick to the Regge--Wheeler (RW) gauge. For practical implementations, after finding a solution in the RW gauge, one should move to an asymptotically flat gauge to ensure regularity of higher order source functions (see \citet{B09}).

The paper is organised as follows: in section \ref{sect:2} we briefly introduce Reissner--Nordstr\"om metric and in section \ref{sect:3} we discuss general form of perturbation expansion of Einstein--Maxwell equations. In sections \ref{sect:4}, \ref{sect:5} and \ref{sect:6} we remind polar expansion in axial symmetry, discuss gauge choice and present source identities. The main result of this paper, namely providing inhomogeneous wave equations for Einstein--Maxwell equations of any perturbation order, is contained in \ref{sect:7}.

\section{Background metric}
\label{sect:2}
Reissner--Nordstr\"om solution describes a static, spherically symmetric black hole with an electric charge. In static coordinates $\left(t\in(-\infty,\infty),r\in(r_+, \infty),\theta\in(0,\pi),\phi\in[0,2\pi)\right)$ it's metric is given by (we use $G=c=4\pi \epsilon_0=1$) :
\eq{
\bar g=-Adt^2+\frac{1}{A}dr^2+r^2d\Omega^2\, , \label{g0}
}
where $A=1-\frac{2M}{r}+\frac{Q^2}{r^2}$, $r_+=M+\sqrt{M^2-Q^2}$, and M and Q are mass and charge of a black hole, respectively (we assume $|Q|<M$). Together with an electromagnetic tensor $\bar F$ with only nonzero terms $\bar F _{tr}=-\bar F _{rt} = \frac{Q}{r}$ metric \eqref{g0} solves Einstein--Maxwell equations:
\eq{
\bar R_{\mu\nu}&=8\pi \bar T_{\mu\nu}\, ,\label{ee1}\\
\bar\nabla_{\mu} \bar F^{\mu\nu}&=0\, ,\label{ee2}\\
\bar F_{[\mu\nu,\lambda]}&=0\, ,\label{ee3}
}
where $\bar \nabla$ and $\bar R_{\mu\nu}$ are, respectively, covariant derivative and Ricci tensor w.r.t.\ metric $\bar g$ and comma denotes partial derivative. $\bar T_{\mu\nu}$ is given by
\eq{
\bar T_{\mu\nu}=\frac{1}{4\pi}(\bar F_{\mu\alpha}\bar F_{\nu}^{\,\,\,\alpha}-\frac{1}{4}\bar g_{\mu\nu}\bar F_{\alpha\beta} \bar F^{\alpha\beta})
}

\section{Gravitational and electromagnetic perturbations of Einstein--Maxwell systems}
\label{sect:3}
Let's assume that metric $\bar g$ and electromagnetic tensor $\bar F$ solve Einstein--Maxwell equations \eqref{ee1}-\eqref{ee3}.
Now we seek for new solutions $g$ and $F$ that we expand around $\bar g$ and $\bar F$ w.r.t.\ to the perturbation parameter $\epsilon$:
\eq{
g_{\mu\nu}=\bar g_{\mu\nu}+\sum\limits_{i>0}{}^{(i)}h_{\mu\nu}\epsilon^i\, ,\label{serg}\\
F_{\mu\nu}=\bar F_{\mu\nu}+\sum\limits_{i>0}{}^{(i)}f_{\mu\nu}\epsilon^i\, . \label{serF}
}
We plug \eqref{serg}, \eqref{serF} into Einstein--Maxwell equations, to obtain their perturbative form of order $i$:
\eq{
\Delta_L{}^{(i)}h_{\mu\nu}-8\pi {}^{(i)}t_{\mu\nu}&={}^{(i)}S^G_{\mu\nu}\, ,\label{pert1}\\
\bar \nabla^\mu {}^{(i)} f_{\mu\nu}-\pert{i}{\Theta_\nu}& ={}^{(i)}S^M_{\nu}\, ,\label{pert2}\\
\pert{i}{f}_{[\mu\nu,\lambda]}~&=0\, ,\label{pert3}
} 
where
 \begin{widetext}
\eq{
 \Delta_L\pert{i}{h}_{\mu\nu}& =\frac{1}{2}(-\bar\nabla^{\alpha}\bar\nabla_{\alpha}\pert{i}{h}_{\mu\nu}-\bar\nabla_\mu\bar\nabla_\nu \pert{i}{h}^{\alpha}_{\,\,\,\alpha}-2\bar R_{\mu\alpha\nu\beta}\pert{i}{h}^{\alpha\beta}+\bar\nabla_\mu\bar\nabla^{\alpha}\pert{i}{h}_{\nu\alpha}+\bar\nabla_{\nu}\bar\nabla^\alpha \pert{i}{h}_{\mu\alpha})\, ,\\
 {\pert{i}{t}}_{\mu\nu}& =2 {\pert{i}{f}}_{\alpha(\mu}\bar F^{\alpha}{}_{\nu)}-\frac{1}{2}{\pert{i}{f}}_{\alpha\beta}\bar F^{\alpha\beta}\bar g_{\mu\nu}+ \left(\frac{1}{2}\bar F_{\alpha\sigma} \bar F_{\beta}{}^{\sigma} \bar g_{\mu\nu}-\bar F_{\mu\alpha}\bar F_{\nu\beta}\right)\pert{i}{h}^{\alpha\beta} -\frac{1}{4}\bar F^2 \pert{i}{h}_{\mu\nu} - \pert{i}{h}_{\alpha (\mu } \bar T^{\alpha }{}_{\nu) }\,,
 }
\end{widetext}
 \eq{
{\pert{i}{\Theta}}_\nu&=\bar g^{\alpha\beta}( \bar F_{\sigma\nu} {}^{(i)}\delta\Gamma^{\sigma}_{\alpha\beta}+\bar F_{\beta\sigma} {}^{(i)}\delta\Gamma^{\sigma}_{\alpha\nu})\, ,\\
{\pert{i}{\delta\Gamma}}^{\sigma}_{\alpha\beta}&=\frac{1}{2}\bar g^{\sigma\delta}\left(-\bar\nabla_\delta {\pert{i}{h}}_{\alpha\beta}+\bar\nabla_\alpha {\pert{i}{h}}_{\beta\delta}+\bar\nabla_\beta {\pert{i}{h}}_{\delta\alpha}\right)\, .
}
Tensor sources ${}^{(i)}S^G_{\mu\nu}$ and vector sources ${}^{(i)}S^M_{\nu}$ are expressed by ${}^{(j<i)} h_{\mu\nu}$ and ${}^{(j<i)} f_{\mu\nu}$, therefore perturbative Einstein equations should be solved order by order (see Appendix \ref{appa} for the construction of sources). For $i=1$ both sources vanish.

\section{Polar expansion}
\label{sect:4}
In a spherically symmetric background, in 3+1 dimensions, vector and tensor components split into two sectors that transform differently under rotations: polar and axial (for the details see e.g.\ \citep{RW57,Z70a,Z70b,HPN99}). Symmetric tensors have 7 polar and 3 axial components, and antisymmetric tensors have 3 polar and 3 axial components. Below we list expansions of all the components of both symmetric and antisymmetric tensors and of vectors in axial symmetry ($P_\ell$ denotes $\ell$-th Legendre polynomial). 

Symmetric tensor, polar sector:
\eq{
&S_{ab}(t,r,\theta)=\sum\limits_{0\leq \ell} {S_\ell}_{ab}(t,r) P_\ell(\cos\theta)\,, \quad a,b=t,r\, ,\\
&S_{a\theta}(t,r,\theta)=\sum\limits_{1\leq \ell} {S_\ell}_{a\theta}(t,r) \partial_\theta P_\ell(\cos\theta)\,, \quad a=t,r\, ,\\
&\frac{1}{2}\left(S_{\theta\theta}(t,r,\theta)+\frac{S_{\phi\phi}(t,r,\theta)}{\sin^2\theta}\right)=\sum\limits_{0\leq \ell} {S_\ell}_{+}(t,r) P_\ell(\cos\theta)\, ,\\
&\frac{1}{2}\left(S_{\theta\theta}(t,r,\theta)-\frac{S_{\phi\phi}(t,r,\theta)}{\sin^2\theta}\right)= \nonumber \\
&=\sum\limits_{2\leq \ell} {S_\ell}_{-}(t,r) (-\ell(\ell+1)P_\ell(\cos\theta)-2\cot\theta\partial_\theta P_\ell(\cos\theta))\, .
}
Symmetric tensor, axial sector:
\eq{
&S_{a\phi}(t,r,\theta)=\sum\limits_{1\leq \ell} {S_\ell}_{a\phi}(t,r) \sin\theta \partial_\theta P_\ell(\cos\theta)\,, \quad a=t,r\, ,\\
&S_{\theta\phi}(t,r,\theta)=\nonumber\\
&=\sum\limits_{2\leq \ell} {S_\ell}_{\theta\phi}(t,r) \left(-\ell (\ell+1) \sin\theta P_\ell(\cos\theta)-2 \cos \theta \partial_\theta P_\ell(\cos\theta)\right) \, .
}

Antisymmetric tensor, polar sector:
\eq{
A_{tr}(t,r,\theta)&=\sum\limits_{0\leq \ell} {A_\ell}_{tr}(t,r) P_\ell(\cos\theta)\, ,\\
A_{a\theta}(t,r,\theta)&=\sum\limits_{1\leq \ell} {A_\ell}_{a\theta}(t,r) \partial_\theta P_\ell(\cos\theta)\,, \quad a=t,r\, .
}

Antisymmetric tensor, axial sector:
\eq{
A_{a\phi}(t,r,\theta)&=\sum\limits_{1\leq \ell} {A_\ell}_{a\phi}(t,r) \sin\theta \partial_\theta P_\ell(\cos\theta)\,, \quad a=t,r\, ,\\
A_{\theta\phi}(t,r,\theta)&=\sum\limits_{0\leq \ell} {A_\ell}_{\theta\phi}(t,r) \sin\theta P_\ell(\cos\theta)\, .
}
Vector, polar sector:
\eq{
V_a(t,r,\theta)&=\sum\limits_{0\leq \ell} {V_\ell}_{a}(t,r) P_\ell(\cos\theta)\,, \quad a=t,r,\\
V_\theta(t,r,\theta)&=\sum\limits_{1\leq \ell} {V_\ell}_{\theta}(t,r) \partial_\theta P_\ell(\cos\theta)\, .
}
Vector, axial sector:
\eq{
V_\phi(t,r,\theta)&=\sum\limits_{1\leq \ell} {V_\ell}_{\phi}(t,r)\sin\theta \partial_\theta P_\ell(\cos\theta)\, .
}

Since the background is spherically symmetric, differential operators acting on $\pert{i}{h_{\mu\nu}}$ and $\pert{i}{f_{\mu\nu}}$ do not mix axial and polar sectors, therefore Einstein--Maxwell equations split into two sectors as well: there are 7 Einstein and 3 Maxwell equations in polar sector, and 3 Einstein and 1 Maxwell equations in axial sector. In our paper we consider separately $\ell\geq2$, $\ell=1$ and $\ell=0$.

\section{Gauge choice}
\label{sect:5}
Under a gauge transformation induced by a vector $X^\mu $, tensors transform as $t_{\mu\nu}\rightarrow t_{\mu\nu}+\mathcal{L}_{X} t_{\mu\nu}$ (see Appendix \ref{appc} for the explicit form of transformations). Throughout the paper we use Regge--Wheeler gauge \cite{RW57}, namely we set ${{}^{(i)}h_\ell}_{tr}$, ${{}^{(i)}h_\ell}_{r\theta}$ and ${{}^{(i)}h_\ell}_{-}$ to zero in polar sector, and ${{}^{(i)}h_l}_{\theta\phi}=0$ in axial sector. When the background quantities $\bar g$ and $\bar F$ fulfil Einstein equations, left hand sides of perturbation equations \eqref{pert1}-\eqref{pert3} of order $i$ do not feel gauge transformations of order $i$, but source functions ${}^{(i)}S^G_{\mu\nu}$ and ${}^{(i)}S^G_{\mu\nu}$ depend on the gauge transformations of order $j < i$ explicitly, so such a formulation is not fully gauge invariant. This, however, is not a problem, sice equations are solved order by order and for the practical implementations one goes to the asymptotically flat gauge before moving to the next order anyway.

\hspace{20cm}

\section{Sources for Einstein--Maxwell equations}
\label{sect:6}
Sources ${}^{(i)}{S_\ell^G}_{\mu\nu}$ and ${}^{(i)}{S^M_\ell}_{\nu}$ are built of ${\pert{j}{h_\ell}}_{\mu\nu}$ and ${\pert{j}{f_\ell}}_{\mu\nu}$ with $j<i$. These sources are not independent, but fulfil five identities:
\eq{
\bar \nabla^\mu {\pert{i}{S^G}}_{\mu\nu}-\frac{1}{2}\bar \nabla_\nu {\pert{i}{S^G}}^{\mu}_{\,\,\,\mu}-2\bar F^\mu_{\,\,\,\nu} {\pert{i}{S^M}}_\mu=0\, ,\label{idE}\\
\bar\nabla^\mu {\pert{i}{S^M}}_{\mu}=0\, ,\label{idM}
}
which come from Bianchi identity and contracted Jacobi identity for tensor $F_{\mu\nu}$. One can check that they hold using \eqref{pert1}-\eqref{pert3} directly. 
Explicit form of identities \eqref{idE}, \eqref{idM} for polar--expanded sources in polar sector reads (we introduce $\tau = \sqrt{(\ell-1)(\ell+2)}$):

\begin{widetext}
\eq{
& \left({A'}+\frac{2 {A}}{r}\right)   { \pert{i}{S^G_\ell}}_{tr}+\frac{2 Q {A}   }{r^2}{ \pert{i}{S^M_\ell}}_{r}+{A}  \partial_r { \pert{i}{S^G_\ell}}_{tr}-\frac{ 1}{2 {A}}\partial_t { \pert{i}{S^G_\ell}}_{tt}-\frac{1}{2} {A}  \partial_t { \pert{i}{S^G_\ell}}_{rr}-\frac{\ell(\ell+1) }{r^2}{ \pert{i}{S^G_\ell}}_{t\theta}-\frac{ 1}{r^2}\partial_t { \pert{i}{S^G_\ell}}_{+}=0\, ,\label{id0}\\
& \left({A'}+\frac{2 {A}}{r}\right) { \pert{i}{S^G_\ell}}_{rr}+\frac{2 Q }{r^2 {A}}{ \pert{i}{S^M_\ell}}_t+\frac{1  }{2 {A}}\partial_r{ \pert{i}{S^G_\ell}}_{tt}+\frac{1}{2} {A} \partial_r { \pert{i}{S^G_\ell}}_{rr} -\frac{1  }{{A}}\partial_t{ \pert{i}{S^G_\ell}}_{tr}-\frac{\ell(\ell+1)  }{r^2}{ \pert{i}{S^G_\ell}}_{r\theta}-\frac{\partial_r  }{r^2}{ \pert{i}{S^G_\ell}}_+=0\, ,\label{id1}\\
& \left({A'}+\frac{2 {A}}{r}\right) { \pert{i}{S^G_\ell}}_{r\theta} +\frac{1 }{2 {A}}{ \pert{i}{S^G_\ell}}_{tt}-\frac{1}{2} {A} { \pert{i}{S^G_\ell}}_{rr} +{A} \partial_r  S_{r\theta} -\frac{1  }{{A}}\partial_t { \pert{i}{S^G_\ell}}_{t\theta}-\frac{\tau^2}{r^2}{ \pert{i}{S^G_\ell}}_-=0\, ,\label{id2}\\
& \left({A'}+\frac{2 {A}}{r}\right) { \pert{i}{S^M_\ell}}_r +{A}  \partial_r { \pert{i}{S^M_\ell}}_r -\frac{1}{{A}}\partial_t { \pert{i}{S^M_\ell}}_t-\frac{\ell(\ell+1)  }{r^2}{ \pert{i}{S^M_\ell}}_\theta=0\, ,\label{idF}
} 
and in axial sector:
\eq{
&\left(A'+\frac{2 A}{r}\right) { \pert{i}{S^G_\ell}}_{r\phi}+A \partial_r { \pert{i}{S^G_\ell}}_{r\phi}-\frac{\partial_t{ \pert{i}{S^G_\ell}}_{t\phi}}{A}-\frac{{\tau^2}{ \pert{i}{S^G_\ell}}_{\theta\phi}}{r^2}=0\, .\label{id4}
}
\end{widetext}

\section{Gravitational and electromagnetic perturbations}
\label{sect:7}
Now we polar--expand equations \eqref{pert1}-\eqref{pert3}:
\eq{
{{}^{(i)} E_\ell}_{\mu\nu}=\Delta_L{}^{(i)}{h_\ell}_{\mu\nu}-8\pi {}^{(i)}{t_\ell}_{\mu\nu}&=\pert{i}{{S^G_\ell}_{\mu\nu}}\, ,\label{pert1l}\\
{{}^{(i)} J_\ell}_{\nu}=\bar \nabla^\mu {}^{(i)} {f_\ell}_{\mu\nu}-{}^{(i)} {\Theta_\ell}_\nu&=\pert{i}{{S^M_\ell}}_{\nu}\, ,\label{pert2l}\\
{{}^{(i)} f_\ell}_{(\mu\nu,\alpha)}&=0 \label{pert3l}.
}

\subsection{Polar sector, $\ell\geq2$}
Firstly, from \eqref{pert3l} we have:
\eq{
{{}^{(i)} f_\ell}_{tr}=\partial_r{{}^{(i)} f_\ell}_{t\theta}-\partial_t{{}^{(i)} f_\ell}_{r\theta}\, ,\label{alg1}
}
and from ${{}^{(i)} E_\ell}_{-}$:
\eq{
\frac{1}{4} \left( \frac{1}{A} \, {}^{(i)}{h}_{\ell\,\,tt}  - A \, {}^{(i)}{h}_{\ell\,\,rr} \right) - {}^{(i)}{S}_{\ell\,\,-} = 0\, .\label{alg2}
}
We use relations \eqref{alg1} and \eqref{alg2} to eliminate ${{}^{(i)} f_\ell}_{tr}$ and ${}^{(i)}{h}_{\ell\,\,tt}$ from equations \eqref{pert1l}--\eqref{pert3l}. Then we are left with 5 variables: ${}^{(i)}{h}_{\ell\,\,tt}$, ${}^{(i)}{h}_{\ell\,\,tr}$, ${}^{(i)}{h}_{\ell\,\,+}$, ${{}^{(i)} f_\ell}_{t\theta}$ and ${{}^{(i)} f_\ell}_{r\theta}$.

Remaining equations can be all fulfilled by introducing two master scalar variables $\PhiP$ and $\PsiP$ which solve a system of two coupled inhomogeneous (homogeneous at the linear order) wave equations
\footnote{In Zerilli's paper coupling potentials to $\PhiP$ and $\PsiP$ were different, and there was an additional coupling potential to the derivative of $\PhiP$ as well. However, there is a linear transformation between Zerilli's and ours master scalar functions.}:
\eq{
r(- \bar \Box + \VPG)\frac{\PhiP}{r}+\VPS\PsiP=\SPG\, ,\label{waveG}\\
r(- \bar \Box + \VPM)\frac{\PsiP}{r}+\VPS\PhiP=\SPM\, .\label{waveM}
}
Following the idea of \cite{AR17}, we express leftover variables by linear combinations of master scalar functions, their partial derivatives up to the second order (to solve homogeneous part of Einstein--Maxwel equations) and additional source functions (to solve inhomogeneous part of equations). These combinations and potentials $\VPG$, $\VPM$, $\VPS$ are defined uniquely:
\begin{widetext}
\eq{
\VPG&=\tau^2 \VPGHAT=\frac{{\tau^2} \left(-r^2 {A'}^2-2 {A} \left(-2 {A}+\ell(\ell+1)+2\right)+\ell^2 (\ell+1)^2\right)}{r^2 \left(r {A'}-2 {A}+\ell(\ell+1)\right)^2}+\frac{8 Q^2 {\tau^2}  {A}}{r^4 \left(r {A'}-2 {A}+\ell(\ell+1)\right)^2}\, ,\label{VPG}\\
\VPM&=\frac{-r {A'}+\ell(\ell+1)}{r^2}+\frac{4 Q^2 \left(2 {A} \left(2 r^3 {A'}+{\tau^2} r^2+4 Q^2\right)-r^4 {A'}^2-4 r^2 {A}^2+\left(\ell(\ell+1)\right)^2 r^2\right)}{r^6 \left(r {A'}-2 {A}+\ell(\ell+1)\right)^2}\, ,\label{VPM}\\
\VPS&=\tau  \VPSHAT=\frac{2 \tau Q \left(2 {A} \left(r^3 {A'}+4 Q^2-2 r^2\right)-r^4 {A'}^2+\left(\ell(\ell+1)\right)^2 r^2\right)}{r^5 \left(r {A'}-2 {A}+\ell(\ell+1)\right)^2}\, ,\label{VPS}\\
\pert{i}{h}_{\ell\,\,tr} = & -r\partial_{tr}\PhiP+  \left(\frac{r {A'}}{2 {A}}-\frac{\tau^2}{r {A'}-2 {A}+\ell(\ell+1)}\right)\partial_t\PhiP-\frac{2 \tau Q \partial_t  }{r \left(r {A'}-2 {A}+\ell(\ell+1)\right)}\PsiP+\anon\, ,\label{polarsol_1}\\
\pert{i}{h}_{\ell\,\,rr} =&-r  \partial_{rr} \PhiP+ \left(-\frac{\tau^2}{r {A'}-2 {A}+\ell(\ell+1)}-\frac{r {A'}}{2 {A}}\right)\partial_r\PhiP +\frac{r}{2A}{ \left(\frac{A'}{r}+\VPG\right)}\PhiP +\nonumber \label{}\\
&-\frac{2 \tau Q  }{r \left(r {A'}-2 {A}+\ell(\ell+1)\right)}\partial_r \PsiP+\frac{r  }{2 {A}}\VPS\PsiP+\bnon\, ,\label{polarsol_2}\\
\pert{i}{h}_{\ell\,\,+} = &-{A} \partial_r \PhiP +\frac{\left(\frac{2 {A} \left(r {A'}-2 {A}+2\right)}{r {A'}-2 {A}+\ell(\ell+1)}-\ell(\ell+1)\right)}{2 r}\PhiP-\frac{2 \tau Q {A} }{r^2 \left(r {A'}-2 {A}+\ell(\ell+1)\right)}\PsiP +\gnon\, ,\label{polarsol_3}\\
\pert{i}{f}_{\ell\,\,t\theta} = &\frac{A \tau}{4} \partial_r \PsiP-\frac{Q A}{2 r} \partial_r \PhiP+\frac{Q {A} }{2 r^2}\PhiP +\lnon \, ,\label{polarsol_4}\\
\pert{i}{f}_{\ell\,\,r\theta} = &\frac{\tau }{4 {A}} \partial_t\PsiP-\frac{Q  }{2 r {A}}\partial_t \PhiP+\knon\, .\label{polarsol_5}
}
At the linear level $\pert{1}\alpha_\ell=\pert{1}\beta_\ell=\pert{1}\gamma_\ell=\pert{1}\lambda_\ell=\pert{1}\kappa_\ell=\pert{1}{\tilde S}_{G\, \ell}^\Psc=\pert{1}{\tilde S}_{M\, \ell}^\Psc=0$ and relations \eqref{polarsol_1}--\eqref{polarsol_2} can be inverted to express $\pert{1}\Phi^{\Psc}_\ell$ and $\pert{1}\Psi^{\Psc}_\ell$ as functions of ${\pert{i}{h}_\ell}_{\mu\nu}$ and ${\pert{i}{f}_\ell}_{\mu\nu}$. At higher orders, we treat linear level expressions for $\pert{1}\Phi^{\Psc}_\ell$ and $\pert{1}\Psi^{\Psc}_\ell$ as definitions of $\PhiP$ and $\PsiP$:

\eq{
\PhiP&=\frac{4 r A \left(r \partial_r{{\pert{i}h_\ell}_+} -A {\pert{i}h_\ell}_{rr}\right)}{\ell(\ell+1) \left(r A'-2 A+\ell(\ell+1)\right)}-\frac{2 r {\pert{i}h_\ell}_+}{\ell(\ell+1)}\, ,\label{pdef1}\\
\PsiP&=\frac{4 r^2 \left(\partial_r {\pert{i}f_\ell}_{t\theta}-\partial_t {\pert{i}f_\ell}_{r\theta}\right)}{\ell(\ell+1) \tau}+\frac{8 Q A \left(r \partial_r {\pert{i}h_\ell}_+ -A {\pert{i}h_\ell}_{rr}\right)}{\ell(\ell+1) \tau \left(r A'-2 A+\ell(\ell+1)\right)}\, .\label{pdef2}
}
\end{widetext}
Having these definitions, we may express left hand side of \eqref{waveG}, \eqref{waveM} as combinations of ${{}^{(i)} E_\ell}_{\mu\nu}$, ${{}^{(i)} J_\ell}_{\nu}$, and their derivatives. Finding these combinations, we use \eqref{pert1l} and \eqref{pert2l} to build sources for wave equations:
\begin{widetext}
\eq{
&\SPG=\\
&-\frac{4 A^2 \left(\tau^2 r^2+4 Q^2\right) {{\pert{i}{S^G_\ell}}_{rr}}}{\ell(\ell+1) r \left(r A'-2 A+\ell(\ell+1)\right)^2}+\frac{4 {{\pert{i}{S^G_\ell}}_{tt}} \left(2 r^3 A'-4 r^2 A+\left(\ell(\ell+1)+2\right) r^2-4 Q^2\right)}{\ell(\ell+1) r \left(r A'-2 A+\ell(\ell+1)\right)^2}+\\
&+\frac{8 A \partial_r{{\pert{i}{S^G_\ell}}_{+}}}{\ell(\ell+1) \left(r A'-2 A+\ell(\ell+1)\right)}+\frac{8 A {{\pert{i}{S^G_\ell}}_{r\theta}}}{r A'-2 A+\ell(\ell+1)}-\frac{4 r \VPG {{\pert{i}{S^G_\ell}}_{+}}}{\ell(\ell+1) \tau^2}+\\
&+\frac{4 {{\pert{i}{S^G_\ell}}_{-}} \left(\frac{Q^2 (8 A)}{r^3 \left(r A'-2 A+\ell(\ell+1)\right)}-A'-r \VPM\right)}{\ell(\ell+1)}-\frac{16 Q {\pert{i}{S^M_\ell}}_{t}}{\ell(\ell+1) \left(r A'-2 A+\ell(\ell+1)\right)}\, ,  \label{SG}\\
&\frac{\ell(\ell+1)}{4}\tau \SPM=\frac{\ell(\ell+1)}{4} \SPMHAT=\\
&r^2 \partial_r {\pert{i}{S^M_\ell}}_{t}-r^2 \partial_t {\pert{i}{S^M_\ell}}_{r}-{\pert{i}{S^M_\ell}}_{t} \left(2 r-\frac{8 Q^2}{r \left(r A'-2 A+\ell(\ell+1)\right)}\right)+\\
&+\frac{8 Q \left(\frac{r^2 \left(r A'-2 A+2\right)}{4}- Q^2\right) {\pert{i}{S^G_\ell}}_{tt}}{r^2 \left(r A'-2 A+\ell(\ell+1)\right)^2}+\frac{8 Q A^2 {\pert{i}{S^G_\ell}}_{rr} \left(\frac{r^2 \left(r A'-2 A+2 \left(\ell(\ell+1)-1\right)\right)}{4}+ Q^2\right)}{r^2 \left(r A'-2 A+\ell(\ell+1)\right)^2}+\\
&+\frac{4 \ell(\ell+1) Q A {\pert{i}{S^G_\ell}}_{r\theta}}{r \left(r A'-2 A+\ell(\ell+1)\right)}+\frac{4 Q A \partial_r{\pert{i}{S^G_\ell}}_{+}}{r \left(r A'-2 A+\ell(\ell+1)\right)}+2 Q A' \partial_r{\pert{i}{S^G_\ell}}_{-}+\\
&-\frac{2 Q {\pert{i}{S^G_\ell}}_{-} \left(A'+r \VPM\right)}{r}+2 Q A \partial_r^2{\pert{i}{S^G_\ell}}_{-}-\frac{2 Q \partial_t^2{\pert{i}{S^G_\ell}}_{-}}{A}-\frac{r \VPS {\pert{i}{S^G_\ell}}_{+}}{\tau } \, .\label{SM}
}
\end{widetext}
$\SPG$ and $\SPM$ are given uniquely up to the source identities \eqref{id0}--\eqref{idF}. We introduced auxiliary quantities $\VPGHAT$, $\VPSHAT$, $\SPMHAT$, which are nonzero (or nonsingular) for $\ell=1$.

At the nonlinear level, part of solution \eqref{polarsol_1}--\eqref{polarsol_5} consisting of master scalars $\PhiP$ and $\PsiP$ and their derivatives fulfils the homogeneous part of Einstein--Maxwell equations \eqref{pert1l}, \eqref{pert2l}, \eqref{pert3l}, whereas part of \eqref{polarsol_1}--\eqref{polarsol_5} consisting of functions $\anon,\,\bnon,\,\gnon,\,\lnon,\,\knon$ is responsible for the inhomogeneous part of Einstein--Maxwell equations. To find these functions, we plug \eqref{polarsol_1}--\eqref{polarsol_5} into equations \eqref{pert1l}, \eqref{pert2l} and into definitions \eqref{pdef1} and \eqref{pdef2} to ensure consistency. Then we solve these equations for $\anon,\,\bnon,\,\gnon,\,\lnon,\,\knon$. These functions, as well as scalar sources for wave equations $\SPG$, $\SPM$, are defined uniquely (up to the source identities \eqref{id0}--\eqref{idF}):

\eq{
\anon&=-\frac{2 r^2 \left(r^2 A^2 {\pert{i}{S^G_\ell}}_{rr}+r^2 {\pert{i}{S^G_\ell}}_{tt}+2 A {\pert{i}{S^G_\ell}}_{+}\right)}{\ell(\ell+1) r^2 \left(r A'-2 A+\ell(\ell+1)\right)}+\nonumber\\
&-\frac{16 Q^2 A {\pert{i}{S^G_\ell}}_{-}}{\ell(\ell+1) r^2 \left(r A'-2 A+\ell(\ell+1)\right)}\, , \\
\bnon&=r \left(\frac{2 r {\pert{i}{S^G_\ell}}_{tr}}{\ell(\ell+1)}+\frac{\partial_t\anon}{A}\right)\, ,\\
\gnon&=\frac{r \partial_r\anon+\anon }{A}-\frac{\anon  \left(r A'+\ell(\ell+1)\right)}{2 A^2}\, ,\\
\knon&=\frac{r^2 {\pert{i}{S^M_\ell}}_{r}}{\ell(\ell+1)}+\frac{2 Q \partial_t{\pert{i}{S^G_\ell}}_{-}}{A \ell(\ell+1)}\, ,\\
\lnon&=\frac{r^2 {\pert{i}{S^M_\ell}}_{t}}{\ell(\ell+1)}+\frac{2 Q A \partial_r{\pert{i}{S^G_\ell}}_{-}}{\ell(\ell+1)}\, .
}

\subsection{Polar sector, $\ell = 1$}
For $\ell=1$, there is no ${S}_{\ell\,\,-}$ coefficient in a symmetric tensor decomposition, therefore we don't have $\pert{i}{h}_{\ell\,\,-}$ metric coefficient and we loose algebraic Einstein equation \eqref{alg2}.
However, since one of the gauge conditions was $\pert{i}{h}_{\ell\,\,-}=0$, we gain additional gauge freedom, which we can use to keep algebraic relation \eqref{alg2}. That means, our $\ell\geq2$ results are directly applicable to $\ell=1$ as well. The only obstacle is that for $\ell=1$ cofficient $\tau=0$ and singular terms appear in the source for wave equation \eqref{waveM} and in the definition \eqref{pdef2}. We can deal with it introducing $\hat\PsiP=\tau \PsiP$, which, together with $\PhiP$, fulfils a set of wave equations:
\eq{
r(- \bar \Box + \tau^2\VPGHAT)\frac{\PhiP}{r}+\VPSHAT\hat \PsiP=\SPG\, ,\label{waveGl1}\\
r(- \bar \Box + \VPM)\frac{\hat \PsiP}{r}+\tau^2\VPSHAT\PhiP= \SPMHAT\, ,\label{waveMl1}
}
where $\VPGHAT$,  $\VPSHAT$ and $\SPMHAT$ are defined in \eqref{VPG}, \eqref{VPS} and \eqref{SM}. For $\ell=1$ the system is simpler - there is no coupling to gravitational master scalar in \eqref{waveMl1}. Now scalar sources for both equations are regular for $\ell=1$. Metric and electromagnetic tensor perturbations are then given by:
\begin{widetext}
\eq{
\pert{i}{h}_{\textit{1}\,\,tr} = & -r\partial_{tr}\PhiPone+  \frac{r {A'}}{2 {A}}\partial_t\PhiPone-\frac{2  Q \partial_t  }{r \left(r {A'}-2 {A}+2\right)}\PsiPHATone+\anonone\, ,\label{polarsol_1l1}\\
\pert{i}{h}_{\textit{1}\,\,rr} =&-r  \partial_{rr} \PhiPone-\frac{r {A'}}{2 {A}}\partial_r\PhiPone +\frac{A'}{2A}{}\PhiPone +\nonumber \label{}\\
&-\frac{2  Q  }{r \left(r {A'}-2 {A}+2\right)}\partial_r \PsiPHATone+\frac{r  }{2 {A}}\VPSHATone\PsiPHATone+\bnonone\, ,\label{polarsol_2l1}\\
\pert{i}{h}_{\textit{1}\,\,+} = &-{A} \partial_r \PhiPone +\frac{A-1}{ r}\PhiPone-\frac{2  Q {A} }{r^2 \left(r {A'}-2 {A}+2\right)}\PsiPHATone +\gnonone\, ,\label{polarsol_3l1}\\
\pert{i}{f}_{\textit{1}\,\,t\theta} = &\frac{A }{4} \partial_r \PsiPHATone-\frac{Q A}{2 r} \partial_r \PhiPone+\frac{Q {A} }{2 r^2}\PhiPone +\lnonone \, ,\label{polarsol_4l1}\\
\pert{i}{f}_{\textit{1}\,\,r\theta} = &\frac{1 }{4 {A}} \partial_t\PsiPHATone-\frac{Q  }{2 r {A}}\partial_t \PhiPone+\knonone\, .\label{polarsol_5l1}
}
\end{widetext}
Since there is no $\pert{i}{{S^G_\ell}_{-}}$ source term, $\anonone,\,\bnonone,\,\gnonone,\,\lnonone,\,\knonone$ for $\ell=1$ are given by:
\eq{
\anonone&=-\frac{ r^2 A^2 {\pert{i}{S^G_\textit{1}}}_{rr}+r^2 {\pert{i}{S^G_\textit{1}}}_{tt}+2 A {\pert{i}{S^G_\textit{1}}}_{+}}{ \left(r A'-2 A+2\right)}\, ,\\
\bnonone&=r^2 {\pert{i}{S^G_\textit{1}}}_{tr}+\frac{r \partial_t\anonone}{A}\, ,\\
\gnonone&=\frac{r \partial_r\anonone+\anonone }{A}-\frac{\anonone  \left(r A'+2\right)}{2 A^2}\, ,\\
\knonone&=\frac{r^2 }{2}{\pert{i}{S^M_\textit{1}}}_{r}\, ,\\
\lnonone&=\frac{r^2}{2} {\pert{i}{S^M_\textit{1}}}_{t}\, .
}

Although direct implementation of previous results provides a general solution to $\ell=1$ equations, it can be misleading: it looks like there are two dynamical variables, whereas there should be only one \citep{B79} (for Schwarzschild case $\ell=1$ gravitational modes are pure gauge \citep{T69}). However, by the following gauge transformation, one can get rid of $\PhiPone$ from \eqref{polarsol_1l1}-\eqref{polarsol_5l1}: 
\eq{
\pert{i}{\zeta}_{\textit{1}\,\, t} =& -\partial_t \pert{i}{\zeta}_{\textit{1}\,\, \theta}\, ,\\
\pert{i}{\zeta}_{\textit{1}\,\, r} =&  \frac{2 \pert{i}{\zeta}_{\textit{1}\,\, \theta}}{r}-\partial_r \pert{i}{\zeta}_{\textit{1}\,\, \theta}\, ,\\
\pert{i}{\zeta}_{\textit{1}\,\, \theta} =& -\frac{r}{2}\PhiPone\, ,
}
and the solution reads:
\begin{widetext}
\eq{
\pert{i}{h}_{\textit{1}\,\,tt} =&  -\frac{2 A^2 Q  }{r \left(r {A'}-2 {A}+2\right)}\partial_r \PsiPHATone-\frac{r  A }{2 }\VPSHATone\PsiPHATone+A^2 \bnonone + r A \SPG \label{}\\
\pert{i}{h}_{\textit{1}\,\,tr} = &  -\frac{2  Q \partial_t  }{r \left(r {A'}-2 {A}+2\right)}\PsiPHATone+\anonone\, ,\label{polarsol_1l1}\\
\pert{i}{h}_{\textit{1}\,\,rr} =& -\frac{2  Q  }{r \left(r {A'}-2 {A}+2\right)}\partial_r \PsiPHATone+\frac{r  }{2 {A}}\VPSHATone\PsiPHATone+\bnonone\, ,\label{polarsol_2l1}\\
\pert{i}{h}_{\textit{1}\,\,+} = &-\frac{2  Q {A} }{ \left(r {A'}-2 {A}+2\right)}\PsiPHATone +r^2 \gnonone\, ,\label{polarsol_3l1}\\
\pert{i}{f}_{\textit{1}\,\,t\theta} = &\frac{A }{4} \partial_r \PsiPHATone +\lnonone \, ,\label{polarsol_4l1}\\
\pert{i}{f}_{\textit{1}\,\,r\theta} = &\frac{1 }{4 {A}} \partial_t\PsiPHATone+\knonone\, .\label{polarsol_5l1}
}
\end{widetext}

The cost of performing this transformation is the loss of algebraic relation \eqref{alg2}. From our results one can also move to a gauge used by some authors (\citep{M75}, \citep{B79}) in which $\pert{i}{h}_{\textit{1}\,\,+}=0$.

\subsection{Polar sector, $\ell = 0$}
In this case we follow \citet{AR17}. Using gauge freedom we set $\pert{i}{h}_{\textit{0}\,\,+}=0$ and $\pert{i}{h}_{\textit{0}\,\,tr=0}$ and leftover nonzero variables are $\pert{i}{h}_{\textit{0}\,\,tt}$, $\pert{i}{h}_{\textit{0}\,\,rr}$ and $\pert{i}{f}_{\textit{0}\,\,tr}$. From $\pert{i}{E_{\textit{0}\,\,\,01}}$, $\pert{i}{E_{\textit{0}\,\,\,00}}+A^2\pert{i}{E_{\textit{0}\,\,\,11}}$ and $\pert{i}{J_{\textit{0}\,\,\,1}}$ (the only independent equations) we have respectively:
\eq{
&\frac{A}{r}\partial_t \pert{i}{h}_{\textit{0}\,\,rr}=\pert{i}{S^G_{\textit{0}\,\,\,tr}} \,,\label{l0eq1}\\
&\frac{A}{r}\partial_r\left(A \pert{i}{h}_{\textit{0}\,\,rr}-\frac{\pert{i}{h}_{\textit{0}\,\,tt}}{A}\right)=\frac{\pert{i}{S^G_{\textit{0}\,\,\,tt}}}{A}+A\pert{i}{S^G_{\textit{0}\,\,\,rr}}\, ,\label{l0eq2}\\
&\partial_t\left(\pert{i}{f}_{\textit{0}\,\,tr}+\frac{Q}{2r^2}(\frac{\pert{i}{h}_{\textit{0}\,\,tt}}{A}-A \pert{i}{h}_{\textit{0}\,\,rr})\right)=-A \pert{i}{S^M_{\textit{0}\,\,\,r}}\, .\label{l0eq3}
}

These equations can be therefore integrated directly, starting from \eqref{l0eq1}.

\subsection{Axial sector, $\ell\geq2$}
Firstly, we use \eqref{pert3l} to obtain:
\eq{
{\pert{i}{f_\ell}}_{t\phi}=-\frac{\partial_t f_{\theta\phi}}{\ell(\ell+1)}\, ,\\
{\pert{i}{f_\ell}}_{r\phi}=-\frac{\partial_r f_{\theta\phi}}{\ell(\ell+1)}\, .
}

We are left with three variables ${\pert{i}h_\ell}_{t\phi}$, ${\pert{i}h_\ell}_{r\phi}$ and ${\pert{i}f_\ell}_{\theta\phi}$. In the same manner as before, we can fulfil equations \eqref{pert1l}-\eqref{pert2l} by introducing two master scalar variables $\PhiA$ and $\PsiA$, which solve a system of two coupled wave equations:
\eq{
r(- \bar \Box + \VAG)\frac{\PhiA}{r}+\VAS\PsiA=\SAG\, ,\label{waveGA}\\
r(- \bar \Box + \VAM)\frac{\PsiA}{r}+\VAS\PhiA=\SAM\, .\label{waveMA}
}

Following the procedure described in the previous section, we find three potentials and express $h_{t\phi}$, $h_{r\phi}$ and $f_{\theta\phi}$ by master scalars and their derivatives:
\eq{
\VAG=&\frac{r^2 \left(A-3 r A'\right)+\left(\tau^2+1\right) r^2-Q^2}{r^4}\, ,\\
\VAM=&\frac{-A'r^3+\ell(\ell+1) r^2+4 Q^2}{r^4}\, ,\\
\VAS=&-\frac{2 \tau Q}{r^3}\, ,\\
{\pert{i}{h_\ell}}_{t\phi}=&  A\partial_{r}(r\PhiA)+\pert{i}{\sigma_\ell}\, ,\label{axialsol_1}\\
{\pert{i}{h_\ell}}_{r\phi}=&  \frac{r }{A}\partial_{t}\PhiA+\pert{i}{\chi_\ell}\, ,\label{axialsol_2}\\
{\pert{i}{f_\ell}}_{\theta\phi}=&\frac{1}{2}   \ell(\ell+1) \tau\PsiA+\pert{i}{\delta_\ell}\, .\label{axialsol_3}
}
Now we invert above relations for linear order and treat the following expressions as definitions of $\PhiA$ and $\PsiA$ at the nonlinear order:
\eq{
\PhiA=&\frac{\left(r \left(\partial_r {\pert{i}{h_\ell}}_{t\phi}-\partial_t {\pert{i}{h_\ell}}_{r\phi}\right)-2 {\pert{i}{h_\ell}}_{t\phi}\right)}{  \ell(\ell+1) {\tau^2} r}+\nonumber\\
&+\frac{4 Q {\pert{i}{f_\ell}}_{\theta\phi}}{  {\tau^2} }\, ,\label{invaxial1}\\
\PsiA=&\frac{2 {\pert{i}{f_\ell}}_{\theta\phi}}{  \tau \ell(\ell+1)}\, .\label{invaxial2}
}

Finally, we find inhomogeneous functions $\pert{i}{\sigma_\ell}$, $\pert{i}{\chi_\ell}$, $\pert{i}{\delta_\ell}$:
\eq{
\pert{i}{\sigma_\ell}=&\frac{2 r^2 }{{\tau^2}}{\pert{i}{S^G_\ell}}_{t\phi} \, ,\\
\pert{i}{\chi_\ell}=&\frac{2 r^2 }{{\tau^2}}{\pert{i}{S^G_\ell}}_{r\phi} \, ,\\
\pert{i}{\delta_\ell}=&0\, ,
}
and scalar sources $\SAG$, $\SAM$:
\eq{
\SAG&=\frac{2 r \left(\partial_r {\pert{i}{S^G_\ell}}_{t\phi}-\partial_t {\pert{i}{S^G_\ell}}_{r\phi}\right)}{  {\tau^2}}\, ,\label{SGA}\\
\SAM&=\frac{2 {\pert{i}{S^M_\ell}}_{\phi}}{  \tau}\, . \label{SMA}
}

\subsection{Axial sector, $\ell=1$}
Since $\pert{i}{h}_{\ell\,\,\theta\phi}$ does not appear for $\ell=1$, we can use gauge freedom to set $\pert{i}{h}_{1\,\,r\phi}=0$. From \eqref{pert3l} we have:
\eq{
\pert{i}{f}_{\textit{1}\,\,t\phi}=-\frac{\partial_t \pert{i}{f}_{\textit{1}\,\,\theta\phi}}{2}\, ,\\
\pert{i}{f}_{\textit{1}\,\,r\phi}=-\frac{\partial_r \pert{i}{f}_{\textit{1}\,\,\theta\phi}}{2}\, .
}
Remaining equations contain $ \pert{i}{h}_{\textit{1}\,\,t\phi}$ and  $ \pert{i}{f}_{\textit{1}\,\,\theta\phi}$ only. From $\pert{i}{E}_{\textit{1}\,\,r\phi}=\pert{i}{S}^G_{\textit{1}\,\,r\phi}$ we find:
\eq{
-\frac{r^2}{2 A}\partial_r\left(\frac{\pert{i}{h}_{\textit{1}\,\,t\phi}}{r^2}\right)-\frac{Q  \pert{i}{f}_{\textit{1}\,\,\theta\phi}}{A r^2}+\eta(r) =\int^t \pert{i}{S}^G_{\textit{1}\,\,r\phi}\, dt'\, ,
} 
where $\eta(r)$ is some function of $r$. It is not arbitrary - from $\pert{i}{E}_{\textit{1}\,\,t\phi}=\pert{i}{S}^G_{\textit{1}\,\,t\phi}$ and source identity \eqref{id4} we find $\eta = \frac{C_1}{ A r^2}$, $C_1$ being an arbitrary constant.

Let's introduce $\PsiAone$ such that $ \pert{i}{f}_{\textit{1}\,\,\theta\phi} = \PsiAone+\frac{4 C_1 Q}{3 r^2 \left(r A'+2 A-2\right)}$. 
From \eqref{pert2l} we find that $\PhiAone$ fulfils an inhomogeneous (homogeneous at the linear level) wave equation:
\eq{
r(- \bar \Box + \VAMone)\frac{\PsiAone}{r}=\SAMone\, ,
}
where:
\eq{
\VAMone&=\frac{ 4 Q^2-r^3A' +2 r^2}{r^4}\, ,\\
\SAMone &= 2 \pert{i}{S}^M_{\textit{1}\,\,\phi}-\frac{4 A Q \int^t \pert{i}{S}^G_{\textit{1}\,\,r\phi} \, dt'}{r^2} \, .
}

We note that at the linear level setting $\PsiAone=0$ corresponds to the linearised Kerr--Newman metric.

\section{Summary}
\label{sect:8}
Nonlinear perturbation theory of Reissner--Nordstr\"om solution was not present in the literature so far and present article fills this gap. Basing on a systematic approach to gravitational perturbations by \citet{AR17}, we have shown that one can fulfil perturbative Einstein--Maxwell equations at any perturbation order by solving two inhomogeneous master wave equations at each sector (cases $\ell=0,1$ needed special treatment). This makes treatment of higher--order perturbations of Reissner--Nordstr\"om clear and would be especially useful for the the numerical purposes. To summarise, a complete order by order algorithm of solving Einstein--Maxwell equations within our formalism would be:
\begin{enumerate}
\item Solve wave equations \eqref{waveG}, \eqref{waveM}, \eqref{waveGA}, \eqref{waveMA} and calculate RW variables and electromagnetic tensor components according to \eqref{polarsol_1}-\eqref{polarsol_5}, \eqref{axialsol_1}-\eqref{axialsol_3},
\item Move to asymptotically flat gauge and calculate sources to Einstein--Maxwell equations (Appendix \ref{appa}),
\item Construct sources to wave equations (equations \eqref{SG}, \eqref{SM}, \eqref{SGA}, \eqref{SMA}) and move to the next order.
\end{enumerate}
Applications of presented calculations possibly include nonlinear studies on \textit{strong censorship conjecture} and on astrophysical systems, where electromagnetic filed is taken into account.

\begin{acknowledgements}
I am grateful to Andrzej Rostworowski for help. This research was supported by the Polish National Science Centre grant no. 2017/26/A/ST2/00530. 
\end{acknowledgements}

\appendix

\section{Sources for Einstein--Maxwell equations}\label{appa}
Let's fix index $i$ and assume that we already know the solution to Einstein--Maxwell equations \eqref{pert1l}--\eqref{pert3l} up to i-th order:
\eq{
\tilde{g}_{\mu\nu}=\sum\limits_{j=1}^{i}\sum\limits_{\ell} {\pert{j}{h_\ell}}_{\mu\nu}\, ,\\
\tilde F_{\mu\nu}=\sum\limits_{j=1}^{i}\sum\limits_{\ell} {\pert{j}{f_\ell}}_{\mu\nu}\, .
}
Using this solution we can calculate Einstein tensor $G_{\mu\nu}(\tilde g)$ and energy--momentum tensor $ T_{\mu\nu}(\tilde g,\tilde F)$. Although these tensors fulfil Einstein--Maxwell equations up to order $i$, they contribute to the $i+1$ (and higher) perturbation equations. Finally, tensor and vector sources of order i+1 are given by:
\eq{
{\pert{i+1}{S^G}}_{\mu\nu}&=[i+1]\left(-G_{\mu\nu}(\tilde{g})+8\pi T_{\mu\nu}(\tilde{g},\tilde{F})\right)\, ,\\
{\pert{i+1}{S^E}}_{\nu}&=[i+1]\left(-\nabla^{\mu}(\tilde{g}_{\alpha\beta})\tilde{F}_{\mu\nu}\right)\, ,
}
where $[k]\left(...\right)$ denotes the k-th order expansion in $\epsilon$ of a given quantity.

Although in most cases expressions for the sources ${\pert{i+1}{S^G}}_{\mu\nu}$ and ${\pert{i+1}{S^G}}_{\mu\nu}$ are complicated, their construction is a purely algebraic task and can be easily performed using computer algebra.

\section{Gauge transformations}\label{appc}
Under a gauge transformation $x^\mu\rightarrow x^\mu+X^\mu $, tensors transform as $t_{\mu\nu}\rightarrow t_{\mu\nu}+\mathcal{L}_{X} t_{\mu\nu}$. For $X^\mu={}^{(i)}\zeta^\mu \epsilon^i$, perturbation functions of order $i$ transform in the following way:
\eq{
{}^{(i)}{h_\ell}_{\mu\nu} \rightarrow& {}^{(i)}{h_\ell}_{\mu\nu}+\mathcal{L}_{{}^{(i)}{\zeta_\ell}}\bar g_{\mu\nu}\, ,\\
{}^{(i)}{f_\ell}_{\mu\nu} \rightarrow& {}^{(i)}{f_\ell}_{\mu\nu}+\mathcal{L}_{{}^{(i)}{\zeta_\ell}}\bar F_{\mu\nu}\, .
}
Explicit form of these transformations in polar sector is the following:
\eq{
{\pert{i}{h_\ell}}_{tt}\rightarrow& {\pert{i}{h_\ell}}_{tt}+2\partial_t {\pert{i}{\zeta_\ell}}_t-A A' {\pert{i}{\zeta_\ell}} _r\, ,\\
{\pert{i}{h_\ell}}_{tr}\rightarrow&{\pert{i}{h_\ell}}_{tr}+ \partial_r{\pert{i}{\zeta_\ell}}_t+\partial_t {\pert{i}{\zeta_\ell}} _r-\frac{ {A'}}{A}{\pert{i}{\zeta_\ell}} _t\, ,\\
{\pert{i}{h_\ell}}_{t\theta}\rightarrow& {\pert{i}{h_\ell}}_{t\theta}+\partial_t{\pert{i}{\zeta_\ell}}_\theta+{\pert{i}{\zeta_\ell}} _t\, ,\\
{\pert{i}{h_\ell}}_{rr}\rightarrow& {\pert{i}{h_\ell}}_{rr}+ 2 \partial_r{\pert{i}{\zeta_\ell}}_\theta+\frac{ {A'}}{A}{\pert{i}{\zeta_\ell}} _r\, ,}
\eq{
{\pert{i}{h_\ell}}_{r\theta}\rightarrow&   {\pert{i}{h_\ell}}_{r\theta}+\partial_r{\pert{i}{\zeta_\ell}}_\theta-\frac{2}{r}{\pert{i}{\zeta_\ell}} _\theta+{\pert{i}{\zeta_\ell}} _r\, ,\\
{\pert{i}{h_\ell}}_{+}\rightarrow& {\pert{i}{h_\ell}}_+ + 2 A \frac{{\pert{i}{\zeta_\ell}} _r}{r}- \ell(\ell+1)\frac{{\pert{i}{\zeta_\ell}} _\theta}{r^2}\, ,\\
{\pert{i}{h_\ell}}_{-}\rightarrow&{\pert{i}{h_\ell}}_-+ {\pert{i}{\zeta_\ell}} _\theta\, ,\\
{\pert{i}{f_\ell}}_{t\theta}\rightarrow&{\pert{i}{f_\ell}}_{t\theta}+\frac{A  Q}{r^2} {\pert{i}{\zeta_\ell}} _r\, ,\\
{\pert{i}{f_\ell}}_{r\theta}\rightarrow& {\pert{i}{f_\ell}}_{r\theta}+\frac{ Q}{A r^2}{\pert{i}{\zeta_\ell}} _t\, ,\\
{\pert{i}{f_\ell}}_{tr}\rightarrow& {\pert{i}{f_\ell}}_{tr}+\frac{ Q}{A r^2}{\pert{i}{\zeta_\ell}} _t\, ,\\
{\pert{i}{f_\ell}}_{tr}\rightarrow&{\pert{i}{f_\ell}}_{tr}+Q\partial_r\left(\frac{A}{r^2}{\pert{i}{\zeta_\ell}}_r\right)-\frac{Q }{r^2 A}\partial_t{\pert{i}{\zeta_\ell}}_t \, .
}

and in axial sector:
\eq{
{\pert{i}{h_\ell}}_{t\phi}\rightarrow& {\pert{i}{h_\ell}}_{t\phi}+\partial_t {\pert{i}{\zeta_\ell}}_\phi\, ,\\
{\pert{i}{h_\ell}}_{r\phi}\rightarrow&{\pert{i}{h_\ell}}_{r\phi}+ \partial_r{\pert{i}{\zeta_\ell}}_\phi-2\frac{ {\pert{i}{\zeta_\ell}}_\phi}{r}\, ,\\
{\pert{i}{h_\ell}}_{\theta\phi}\rightarrow& {\pert{i}{h_\ell}}_{\theta\phi}+{\pert{i}{\zeta_\ell}}_\phi\, .\\
{\pert{i}{f_\ell}}_{t\phi}\rightarrow& {\pert{i}{f_\ell}}_{t\phi}\, ,\\
{\pert{i}{f_\ell}}_{r\phi}\rightarrow& {\pert{i}{f_\ell}}_{r\phi}\, ,\\
{\pert{i}{f_\ell}}_{\theta\phi}\rightarrow& {\pert{i}{f_\ell}}_{\theta\phi}\, .
}

\bibliography{Maxwell_2_col}
\bibliographystyle{apsrev4-1}

\end{document}